# Offset-free Measurement of Dipolar Couplings in a Single Crystal and Determination of Molecular Orientation

S. Jayanthi[1] and K. V. Ramanathan[2]

[1]Department of Physics, Indian Institute of Science, Bangalore 560012, INDIA

[2]NMR Research Center, Indian Institute of Science, Bangalore 560012, INDIA

## Abstract

Dipolar couplings are an important source of structure as they provide site specific dipolar splittings for aligned samples and hence are extensively used for the study of membrane proteins in lipid bilayers, liquid crystals and single crystals. Of the many Separated Local Field (SLF) techniques existing to avail this information for static oriented systems, PISEMA (Polarization Inversion Spin Exchange at Magic Angle) has found to have wide application due to its many favorable characteristics. However the pulse sequence suffers from its inherent sensitivity to proton resonance frequency offset. We have recently proposed a sequence named DAPT (Dipolar Assisted Polarization Transfer: *S. Jayanthi et al. Chem. Phys. Lett. **439**, 407, 2007.*) for dipolar coupling measurement which is found to be insensitive to proton offsets over wide range. In this presentation, we report the first implementation of the sequence on rigid systems. Experiments were done on a single crystal of N-Acetyl DL-Valine (NAV). Dipolar couplings measured from the 2D DAPT spectrum is used for identifying the orientations of the magnetically in-equivalent molecules in the unit cell. SLF spectrum of NAV is complicated due to the presence of two magnetically in-equivalent molecules in the unit cell and with pairs of splitting for each $C^\alpha$ - $^1H$ and $C^\beta$ - $^1H$. The molecular orientation has been identified by incorporating the crystal coordinates and constructing a three dimensional rotation matrix about an arbitrary axis using Rodrigues' rotation formula with the aid of a MATLAB program. Dipolar couplings measured from the spectrum is used for all the calculations. The algorithm proposed here is simple and hence can be easily accessible for various other studies aimed for structural elucidation.

[1]jayanthi@physics.iisc.ernet.in, mssjayanti@gmail.com

[2]kvr@sif.iisc.ernet.in

# Introduction

Dipolar couplings obtained using solid state NMR methods from static oriented samples have proved to be extremely useful for elucidating protein structures at atomic resolution, studying interactions of peptides and proteins with membranes, deriving the structure and topology of membrane proteins in lipid bilayers and for the study of dynamics and order in liquid crystalline materials [1-5]. The commonly employed strategy for measuring dipolar couplings is the Separated Local Field (SLF) technique which provides site specific information as a two dimensional plot between chemical shifts of the nucleus such as $^{15}$N or $^{13}$C and the heteronuclear dipolar coupling of the nucleus to the neighboring protons [6-9]. Of the various techniques available for carrying out the SLF experiment, PISEMA has acquired considerable popularity and several investigations have been carried out employing PISEMA [10] for the study of static oriented systems. The advantage of PISEMA is its large scaling factor and relatively smaller line-width. However PISEMA has a major shortcoming in that the measured dipolar couplings are sensitive to positioning of the proton carrier frequency and proton chemical shifts. Usually a couple of PISEMA experiments are done with various proton offsets ranging from aliphatic to aromatic to deduce reliable structure [11].

Recently we have proposed a new heteronuclear polarization transfer sequence named DAPT (Dipolar Assisted Polarization Transfer) [12] and demonstrated it as a local field technique in liquid crystals for deducing dipolar splittings [13]. The sequence is also found to be insensitive to wide range of proton chemical shift evolution. The present work reports the first implementation of DAPT on a rigid system like a single crystal of N-Acetly-DL-Valine (NAV) correlating $^{13}$C chemical shift to $^{13}$C-$^{1}$H dipolar couplings. The SLF spectrum of NAV is complicated due to the presence of two magnetically unique molecules in the unit cell [14]. This results in pairs of splittings for each $C^{\alpha}$ - $^{1}$H and $C^{\beta}$ - $^{1}$H and hence one has to perform specific experiments for its identification [15,16]. We present here a simple approach to identify the molecules using the dipolar coupling values measured from the 2D DAPT spectrum.

Towards this a new algorithm is developed that incorporates the crystal co-ordinates of the molecule together with the experimentally measured dipolar couplings.

Subsequently series of rotations are performed to identify $C^{\alpha}$ - $^1H$ to its bonded $C^{\beta}$ - $^1H$. Rotation of the crystal coordinates were done with a matrix generated using Rodrigues' rotation formula [17, 18]. This gives an efficient method for computing the rotation matrix by an angle 'θ' about a fixed axis specified by a unit vector, $u_i$, where {i=x,y,z}[16]. User interactive MATLAB codes were written based on the algorithm. The protocol proposed here is simple and can be extended for secondary structure determination of oriented proteins and peptides.

## Pulse Sequence

DAPT-I [12,13] sequence (**Fig.1a**) comprises of a 90° pulse on the S (rare spin) channel sandwiched between a homonuclear decoupling sequence [19] (BLEW-12 in this case) $P_0$ and $P_{90}$ for a time duration of $\tau_1$ and $\tau_2$ respectively in the I (abundant) channel. The suffix '0' and the '90' represent the relative phase of BLEW-12 pulses in the respective blocks. The average Hamiltonian's corresponding to the heteronuclear dipolar coupling for the $P_0$ and the $P_{90}$ BLEW-12 blocks are given by $\overline{H}_{(1)}^{0}$ and $\overline{H}_{(2)}^{0}$ respectively as

$$\overline{H}_{(1)}^{0} = D_{IS} \frac{4}{3\pi}(2I_X + I_Z)S_Z \qquad (1)$$

$$\overline{H}_{(2)}^{0} = D_{IS} \frac{4}{3\pi}(2I_Y + I_Z)S_Z \qquad (2)$$

$I_Z$ magnetization evolves under the average Hamiltonian for the $P_0$ BLEW-12 sequence and creates anti-phase magnetization terms. The 90° pulse on the S spin along with the evolution under the Hamiltonian $\overline{H}_{(2)}^{0}$ that corresponds to the $P_{90}$ BLEW-12 pulses converts the two spin order term to pure in-phase $S_Y$ magnetization. At the end of the $\tau_2$ period, S spin signals are acquired under heteronuclear spin decoupling. It was shown earlier that this sequence is able to achieve polarization transfer between $^1H$ and $^{13}C$ in static oriented systems [13]. The role of the BLEW-12 pulses in the sequence is to remove homonuclear dipolar couplings among I spins and to create the two spin order term as in INEPT [20], so that efficient polarization transfer between I and S spins can be achieved. A detailed density matrix evolution of the sequence can be found in our earlier publications [12]. Subsequently we have demonstrated the utility of this sequence as a proton evolved local field (PELF) technique, **Fig.1b**, in extracting heteronuclear dipolar couplings in liquid crystals. Simultaneous $180^o$ pulse applied during the $t_1$ evolution refocuses the proton chemical shifts during the dipolar coupling evolution. A compensation pulse 'α', is applied prior to $P_0$ and $P_{90}$ for getting maximum sensitivity. The measured dipolar splittings were further utilized for the calculation of local order parameter and dynamics.

In this article, we have implemented DAPT on a single crystal of NAV correlating $^1$H - $^{13}$C couplings to $^{13}$C chemical shift. This is also the first demonstration of DAPT on rigid systems. The variation of dipolar couplings with respect to proton offsets are monitored systematically. The dipolar splittings from the spectra are subsequently used for identifying the molecules in the unit cell by categorizing the $C^{\alpha}$ - $^1$H to the corresponding $C^{\beta}$ - $^1$H. The protocol followed is explained in detail in the following sections.

## Experimental

All Experiments were done in a Bruker - Avance II - 500 MHz spectrometer with a 5mm double resonance static probe with solenoid r.f. coil. NAV single crystal has been grown from slow evaporation of its aqueous solution. $90^o$ pulse length for proton and carbon was 2.5 and 3.0 μsec respectively. 1D $^{13}$C spectrum was recorded using 1D DAPT sequence shown in **Fig.1a** with $\tau_1 = \tau_2 = 60$ μsec. This corresponds to two cycles of BLEW-12. SPINAL-128 [21] heteronuclear dipolar decoupling was employed during $^{13}$C acquisition. Pre-scan delay was kept at 5 sec. During 2D, the $\tau_2$ time period of DAPT was retained at 60 μsec while $\tau_1$ was evolved as one BLEW-12 per $t_1$. The maximum $t_1$ evolution time was kept at ~ 5msec. 1D and 2D was acquired with 512 and 128 scans respectively. 2D spectra was acquired in magnitude mode and processed after incorporating the scaling factor of 0.475 in the $F_1$.

## The Algorithm

The unit cell of NAV visualized with the aid of 'MERCURY' is shown in **Fig.2a.** The unit cell comprises of four molecules that are symmetry related, out of which two are magnetically unique and hence results in pairs of lines in the spectrum. A single NAV molecule showing $C^{\alpha}$ - $^1H$ and $C^{\beta}$ - $^1H$ is shown in **Fig.2b.** The algorithm is aimed for identifying the $C^{\alpha}$ - $^1H$ to the bonded $C^{\beta}$ - $^1H$ of a chosen molecule. This in other words identifies the magnetically in-equivalent molecules in the unit cell.

1D $^{13}C$ and 2D local field spectra of NAV recorded using DAPT sequence are shown in **Fig. 3(a,b)** respectively. **Fig.3c** shows the variation of the measured dipolar couplings of an arbitrarily selected peak of NAV with respect to proton offset under PISEMA and DAPT. Since the coupling observed from DAPT is insensitive to proton chemical shift evolution, the values measured from the 2D spectra may be straightaway used for structure calibration whereas PISEMA needs to be repeated for various proton offset conditions. The dipolar splittings marked as (I) and (II) in the 2D corresponds to the $C^{\alpha}$ - $^1H$ couplings from the two molecules. They exist are asymmetric triplets because of the presence of $^{14}N$ in its neighbourhood.

The task is to identify $C^{\beta}$ - $^1H$ corresponding to molecules (I) and (II) from the spectra. The developed algorithm is based on the well know equation **(Eqn.3)** that relates dipolar couplings ($D_{IS}$) between I and S to 'θ', the orientation of the dipolar vector with respect to the applied field $B_o$, and '$r_{IS}$', the bond distance. Since $D_{IS}$ is proportional to ($3\cos^2\theta$-1), it varies sinusoidally with respect to 'θ' (**Fig.4.**) Since $D_{IS}$ transverse through positive and negative values of 'θ' both positive and negative values of $D_{IS}$ are considered.

$$\pm D_{IS} = -\frac{\mu_0 \gamma_I \gamma_S \hbar}{4\pi r_{IS}^3} \frac{1}{2}(3\cos^2\theta - 1) \qquad (3)$$

The protocol is as follows. From the 2D, the dipolar coupling of any one of the $C^{\alpha}-^1H$ is measured, say $D_{IS}{}^{\alpha}(I)$. Similarly from the crystal coordinate file, any one of the molecules is identified as (I) and the corresponding '$r_{IS}$' is calculated or measured using the visualization software MERCURY. Using the above equation, the two possible angles, $\theta^{\alpha}_{+}(I)$ and $\theta^{\alpha}_{-}(I)$, are calculated. The magnetic field direction has to be fixed on the selected molecule for further calculations. Towards this an arbitrary coordinate system is assigned with $C^{\alpha}(I)$ as origin (0, 0, 0) and fixing $B_0$ at (0, 0, 1). Subsequently, the prior $C^{\alpha}(I)$ coordinate is subtracted from the coordinate of entire atoms of all the four molecules in the unit cell. This provides a translational displacement to the unit cell as a whole to the new coordinate system. Having fixed $B_0$, a vector $C^{\alpha}$-A is constructed perpendicular to the vectors $C^{\alpha}-^1H$ (I) and $C^{\alpha}(I)$-$B_0$ which acts as an arbitrary axis of rotation. Using $C^{\alpha}$-A, Rodrigues' rotation matrix is applied to rotate $C^{\alpha}-^1H$ (I) vector by an angle $\theta^{\alpha}_{+}(I)$ initially. The form of the rotation matrix R is given below.

$$R = \begin{bmatrix} u_x^2 + (1-u_x^2)\cos\theta & u_x u_y(1-\cos\theta) - u_z\sin\theta & u_x u_z(1-\cos\theta) + u_y\sin\theta \\ u_x u_y(1-\cos\theta) + u_z\sin\theta & u_y^2 + (1-u_y^2)\cos\theta & u_y u_z(1-\cos\theta) - u_x\sin\theta \\ u_x u_z(1-\cos\theta) - u_y\sin\theta & u_y u_z(1-\cos\theta) + u_x\sin\theta & u_z^2 + (1-u_z^2)\cos\theta \end{bmatrix} \quad (4)$$

where $u_i$, {i=x,y,z}, represents the normalized component of the perpendicular vector $C^{\alpha}$-A. $u_i$ also satisfies the condition $u_x^2 + u_y^2 + u_z^2 = 1$. $\theta$ in Eqn.(4) represents $\theta^{\alpha}_{+}(I)$ initially. This rotation is done to position the $C^{\alpha}-^1H$ (I) dipolar vector of the first molecule (I) by an angle $\theta^{\alpha}_{+}(I)$ with respect to $B_0$. The rotation fixes the molecule (I) hence satisfying the experimental conditions. The outcome of this is an overall rotation of the molecule (I) and hence results in a new orientation of the dipolar vector $C^{\beta}-^1H$ (I) with respect to $B_0$. The new coordinates (N') of $C^{\beta}-^1H$ (I) are calculated through a rotation of the old coordinates (O) by R (N'=R.O). The new orientation of $C^{\beta}-^1H$ (I), say $\theta^{\beta}_{+}(I)$ , with respect to $B_0$ is calculated by constructing a vector $C^{\beta}$ - $B_0$ parallel to $C^{\alpha}$ - $B_0$ at $C^{\beta}(I)$. This can be achieved by a coordinate translation of $B_0$ at $C^{\beta}$. From the value of the angle $\theta^{\beta}_{+}(I)$, the dipolar coupling, $D_{IS}{}^{\beta}(I)$ is calculated using **Eqn.(3).** The calculated value is subsequently compared with the experimental plot for an accurate match. If the results found matching that means a successful identification of $(C^{\alpha} - {}^{1}H)$, $(C^{\beta} - {}^{1}H)$ pair of the same molecule, (I) in this case. Otherwise the whole process has to be repeated for the negative value of $\theta^{\alpha}$.(I) to subsequently end up in $\theta^{\beta}$.(I) and a new value of $D_{IS}{}^{\beta}$(I).

Having obtained $D_{IS}{}^{\beta}$(I), the whole protocol is repeated for an arbitrarily chosen molecule (II) with the same orientation to calculate its $D_{IS}{}^{\alpha}$(II) and $D_{IS}{}^{\beta}$(II). It has to be noted that since ${}^{14}N$ is not in the neighbourhood of $C^{\beta} - {}^{1}H$, the corresponding cross peak is a single dipolar doublet separated by $2D_{IS}{}^{\beta}$.

The above mentioned protocol can be repeated by choosing molecule (II) as origin and calculating angles and the respective couplings for molecule (I). The values obtained for one such case ( (I) as origin) that identifies the molecular orientation is tabulated in **Table-I**. A schematic of the above mentioned flowchart is given in **Fig.5**. A user interactive MATLAB program is written to facilitate the whole procedure mentioned above.

## Results and discussions

NAV is a popular model peptide for solid state NMR studies as it has one peptide linkage. The crystal structure of NAV shows the presence of two magnetically unique molecules per unit cell. This has resulted in doubling of the resonances and hence difficult to identify the $C^{\alpha} - {}^{1}H$ and the respective $C^{\beta} - {}^{1}H$ of the same molecule. ${}^{13}C$ 1D spectrum recorded using DAPT sequence (**Fig1a)** is shown in **Fig.3a**. The 1D is recorded with 512 scans. $\tau_1$ and $\tau_2$ was kept at 60 $\mu$sec. This is also the first demonstration of DAPT on a rigid system like single crystal. The signal to noise obtained was satisfactory to perform further 2D experiments. The major feature of the 1D spectrum is the pairing of resonances especially visible for the carbonyl's. The asymmetric triplets identified as (a) - (d) represents the four carbonyl carbons arising from the two molecules of NAV. This has been identified by using the spectral editing technique, cross polarization polarization inversion (CPPI: data not shown) [22]. The asymmetric splittings are due to the presence of the abundant spin-1 nuclei, ${}^{14}N$.

The aliphatic resonances are grouped within the range of 20-30 ppm hence difficult to identify individually. The 2D spectra recorded using DAPT is shown in **Fig.3b**. The asymmetric triplets from the two $C^\alpha$ - $^1H$ molecules are resolved here with respect to their dipolar couplings and are arbitrarly labeled as (I) and (II) that corresponds to the $^{13}C$ chemical shift span of (41-47 ppm) and (45-60 ppm) respectively. Yet another noticeable feature of the 2D is the presence of tripet arising for the carbonyls (a, d) due to long range couplings which were absent for (b ,c). This we attribute to the spatial proximity of carbonyl's in the molecule. Moreover their visibility in the spectrum is attributed to the nature of the evolution of the couplings explained elsewhere [23-25].

The measured 8.1 and 9.9 kHz are arbitrarily assigned for the molecuels (I) and (II) respectively. Since it is impossible to assign a value nor identify the two $C^\beta$ - $^1H$ dipolar splittings from the spectra, we have employed numerical methods for identification. Following the above mentioned algorithm, we have initially considered $D_{IS}^\alpha(I)$ as +8.1 kHz. The angle $\theta^\alpha_+(I)$ obtained after following the above mentioned algorithm is $66.1^o$. Subsequently, the measured re-orientation angle of $C^\beta$ - $^1H$ with respect to the field $B_0$ is $66.5^o$. This corresponds to a $D_{IS}^\beta(I)$ value of 7.5 kHz. For an arbitrarily chosen molecule-II, in the new coordinate system, $\theta^\alpha_+(II)$ is calculated as $90.8^o$ which gives a dipolar coupling value $D_{IS}^\alpha(II)$ as 15.9 kHz. The corresponding $\theta^\beta_+(II)$ and $D_{IS}^\beta(II)$ are $155.5^o$ and -21.3 kHz respectively. Since the $F_1$ spectral width is kept only at 15 kHz, a dipolar coupling value of $\pm21.3$ kHz may appear after folding at $\pm6.3$ kHz. Since there are no contours in the 2D around 6-8 kHz it can be concluded that the dipolar coupling corresponding to the asymmetric triplets at 8.1 kHz may not be positive.

Subsequently the calculations are repeated for negative $D_{IS}^\alpha(I)$ value, i.e. -8.1 kHz. The calculated values for $D_{IS}^\beta(I)$ is 4.5 kHz for a $\theta^\beta_-(I)$ of $61.4^o$. The corresponding values for molecule-II in the new coordinate system are -27 kHz and $168.5^o$ respectively. As considered above ± 27 kHz can appear after folding at ±12 kHz. Moreover a dipolar cross peak is observed at ~3.5 and 12 kHz in the 2D spectrum. To reconfirm the orientation we have further calculated $D_{IS}^\alpha(II)$ which is found to be 10.8 kHz. Considering an error of ±1 kHz, it can be confirmed that if the origin is fixed on $C^\alpha$ - $^1H$ (I) of value 8.1 kHz, then its corresponding $C^\beta$ - $^1H$ (I) will appear at ~ 3.5 kHz. For the second molecules this provides a $C^\alpha$ - $^1H$ (II) of ~ 10 kHz with its corresponding $C^\beta$ - $^1H$ (II) at 12 kHz. The observations are tabulated in **Table-I** and related pairs are marked in the 2D spectra. We have also repeated the whole process for molecule III and IV to check the reproducibility of the values and is shown in **Table-I**. It is found that the arbitrarily chosen molecules (I − IV) and (II-III) are symmetry related as they provided identical values of angles and dipolar couplings (Table-I). The whole calculation can be repeated by keeping the origin at II molecule and transforming and rotating (I) (data not shown). It has been observed that the resultant values are not matching with the experimental and hence discarded. Furthermore one can also perform the series of calculations with ±9.9 kHz dipolar coupling value and assigning it to any one of the molecule chosen arbitrarily.

## Conclusions

A new algorithm to identify molecular orientations in a single crystal of NAV is presented. NAV has two magnetically in-equivalent molecules and hence results in pairing of resonances therefore impossible to identify the molecular orientation's using a single correlation. The algorithm utilizes the heteronuclear dipolar coupling values measured using a recently proposed local field technique, DAPT. This is the also the first report of DAPT on a rigid system. The experiment is found to be performing well similar to liquid crystalline systems as reported earlier. The experimental values are free from proton offset evolution and hence are reliable for structural calculation. In the case of single crystal though there is a reduction in signal to noise, the recorded spectra was good enough for measuring dipolar couplings without ambiguity. The identification of the molecule is done with the aid of PDB coordinates of NAV and constructing a transformation matrix using Rodrigues' formula by arbitrary fixing the magnetic field on $C^\alpha$ - $^1H$ of any one of the molecule. The dipolar vector of the selected molecule was rotated by the rotation matrix so as to match the experimental plot. This defines a unique orientation to the unit cell and the respective angles and dipolar couplings for $C^\alpha$ - $^1H$ and $C^\beta$ - $^1H$ for the remaining molecules are calculated. The calculated dipolar splittings are then compared with the 2D spectra. This has provided the molecular orientation of $C^\alpha$ -

$^1$H to the corresponding C$^\beta$ - $^1$H of the same molecule. We have also identified the molecules that are symmetry related using this algorithm.

Though there may be quite a few algorithms in the literature towards calculation of secondary structure, we found that the above mentioned work is simple and can be implemented using PDB coordinates and MATLAB codes. This algorithm is not particular with DAPT and hence can be associated with many of the recently proposed offset compensated pulse sequences [26-28] in retrieving dipolar couplings towards determining orientation of systems such as membrane proteins and peptides oriented in lipid bi-layers and single crystals.

**Acknowledgements**

The authors gratefully acknowledge Prof. T. N. Guru Row and Dr. Susanta K Nayak, Solid State Structural Chemistry Unit – Indian Institute of Science – Bangalore-12, INDIA for providing us information regarding crystal structure and visualization software's. Mr. Y. J. S. Reddy, Department of Physics and NMR Research Centre, Indian Institute of Science – Bangalore-12, INDIA is acknowledged for his involvement in the numerical calculations.

**Figure Captions**

Fig.1. (a) DAPT-I (b) DAPT-III pulse sequences. $P_0$ and $P_{90}$ represent BLEW-12 blocks for a time period $\tau_1$ and $\tau_2$ respectively. $P_{90}$ denotes BLEW-12 pulses with a phase difference of $90^o$ when compared with $P_0$. Compensation pulses, $(\alpha)$ with respective phases are applied prior to $P_0$ and $P_{90}$ in (b) to take care of sensitivity loss [7].

Fig.2. (a) Unit cell of the single crystal N-Acetly-DL-Valine (NAV) visualized using the software ''MERCURY'. The four molecules are symmetry related out of which two are magnetically unique. (b) An NAV molecule with $C^\alpha$ - $^1H$ and $C^\beta$ - $^1H$ labelled.

Fig.3. NAV spectrum (a) 1D $^{13}C$ spectrum recorded using the pulse sequence shown in Fig.1a. (b) 2D DAPT spectrum correlating the dipolar coupling between $^1H$ and $^{13}C$ in the $F_1$ to $^{13}C$ chemical shift in the $F_2$. 2D spectrum is recorded using the pulse sequence shown in Fig.1b by keeping $\tau_2$ constant and incrementing $\tau_1$ as $t_1$. (c) Plot showing the variation of the measured dipolar coupling with respect to proton offset for the $^{13}C$ peak at ~ 40ppm for PISEMA and DAPT is shown. The variation for PISEMA clearly depicts the quadratic nature of the measured couplings with proton offset.

Fig.4. Plot showing the variation of the angle 'θ' (between the dipolar vector and the applied magnetic field $B_o$) with respect to the dipolar coupling $(3cos^2\theta-1)$.

Fig.5. Schematic of the algorithm that is explained in the text.

# TABLE-I

Table shows a comparison between the measured and the numerically calculated dipolar coupling values for arbitrarily chosen molecules (I-IV) of NAV single crystal. Molecule (I) is arbitrarily chosen and any one of the dipolar coupling value from the 2D spectrum, say ±8.1 kHz is assigned to that. The orientation and the couplings are measured for molecules II-IV. From the table it can be seen that the $C^\alpha$ - $^1H$ and its bonded $C^\beta$ - $^1H$ pairs are (-8.1, 4.5) kHz and (10.8, 12) kHz respectively. The comparison is done within an error limit of ±1 kHz. Further it has to be noticed that the molecule I,IV and II,III are magnetically equivalent.

| Sign of dipolar coupling | Molecule-I | | | | Molecule -II | | |
|---|---|---|---|---|---|---|---|
| | ( From DAPT) Measured $C^\alpha$ - $^1H$ (kHz) | $\theta^\alpha(I)^o$ | $\theta^\beta(I)^o$ | Calculated $C^\beta$ - $^1H$ (I) (kHz) | $\theta^\beta(II)^o$ | Calculated $C^\beta$ - $^1H$ (II) (kHz) | Calculated $C^\alpha$ - $^1H$ (II) (kHz) |
| Positive | +8.1 | 66.1 | 66.5 | 7.5 | 155.5$^o$ | -21.3 (= 6.3) | 15.8 |
| Negative | **-8.1** | 44.8 | 1.4 | **4.5** | 168.5 | **-27 (=12)** | **10.8** |

| Orientation Fixed wrt to (-8.1 kHz ) of molecule-I | Molecule-III | | | Molecule-IV | | |
|---|---|---|---|---|---|---|
| | $\theta^\beta(III)^o$ | Calculated $C^\beta$ - $^1H$ (III) (kHz) | Calculated $C^\alpha$ - $^1H$ (III) (kHz) | $\theta^\beta(IV)^o$ | Calculated $C^\beta$ - $^1H$ (IV) (kHz) | Calculated $C^\alpha$ - $^1H$ (IV) (kHz) |
| | 11.5 | **-27 (=12)** | **10.8** | 118.5 | **4.5** | **-8.1** |

**Fig.1**

**(a)**

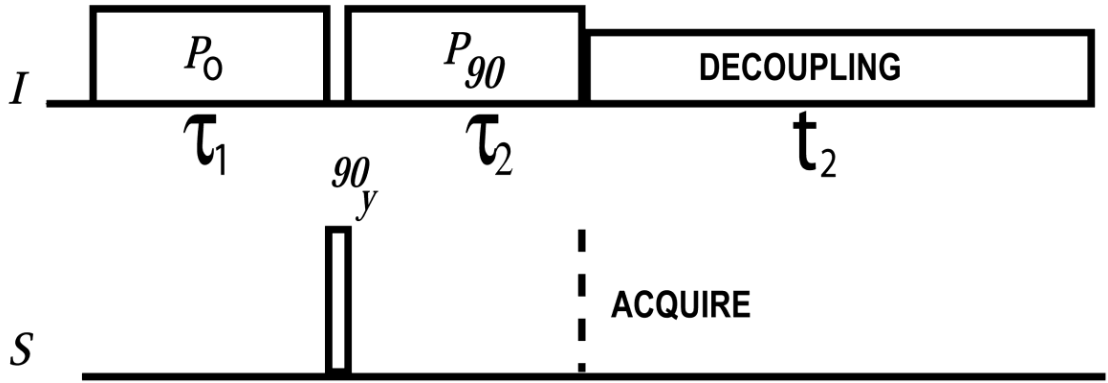

**(b)**

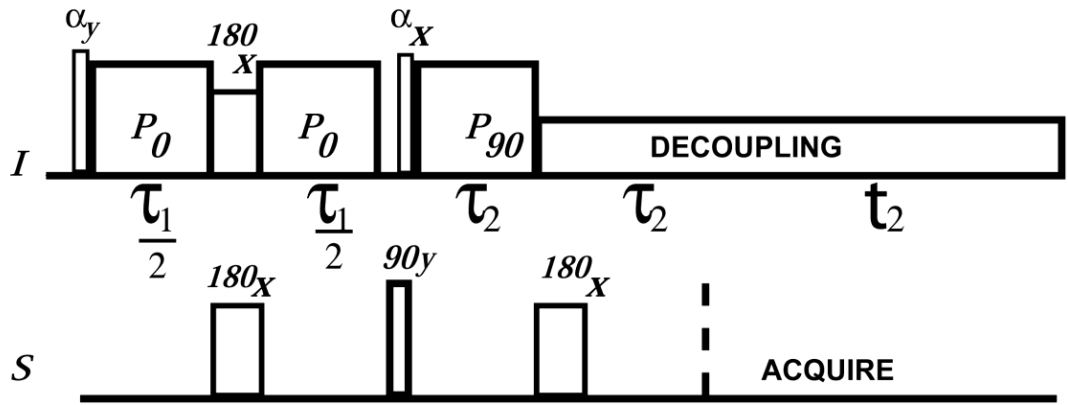

**Fig. 2**

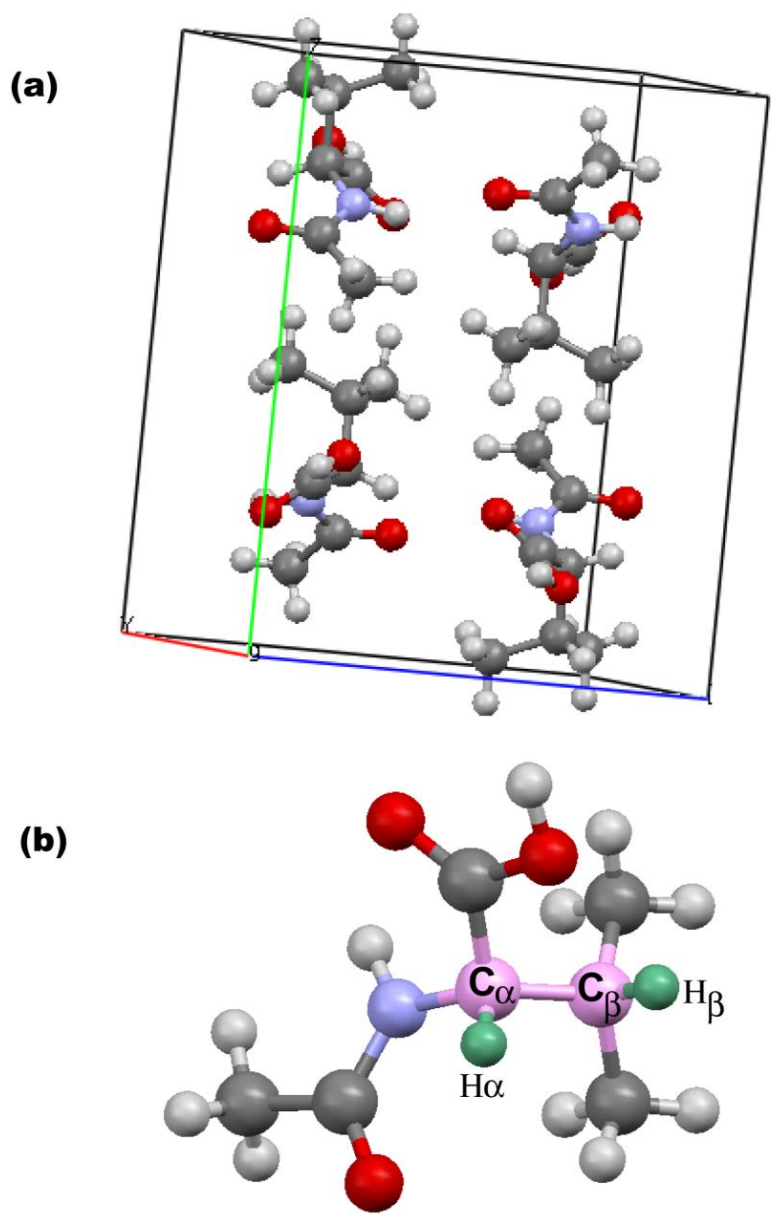

**Fig.3**

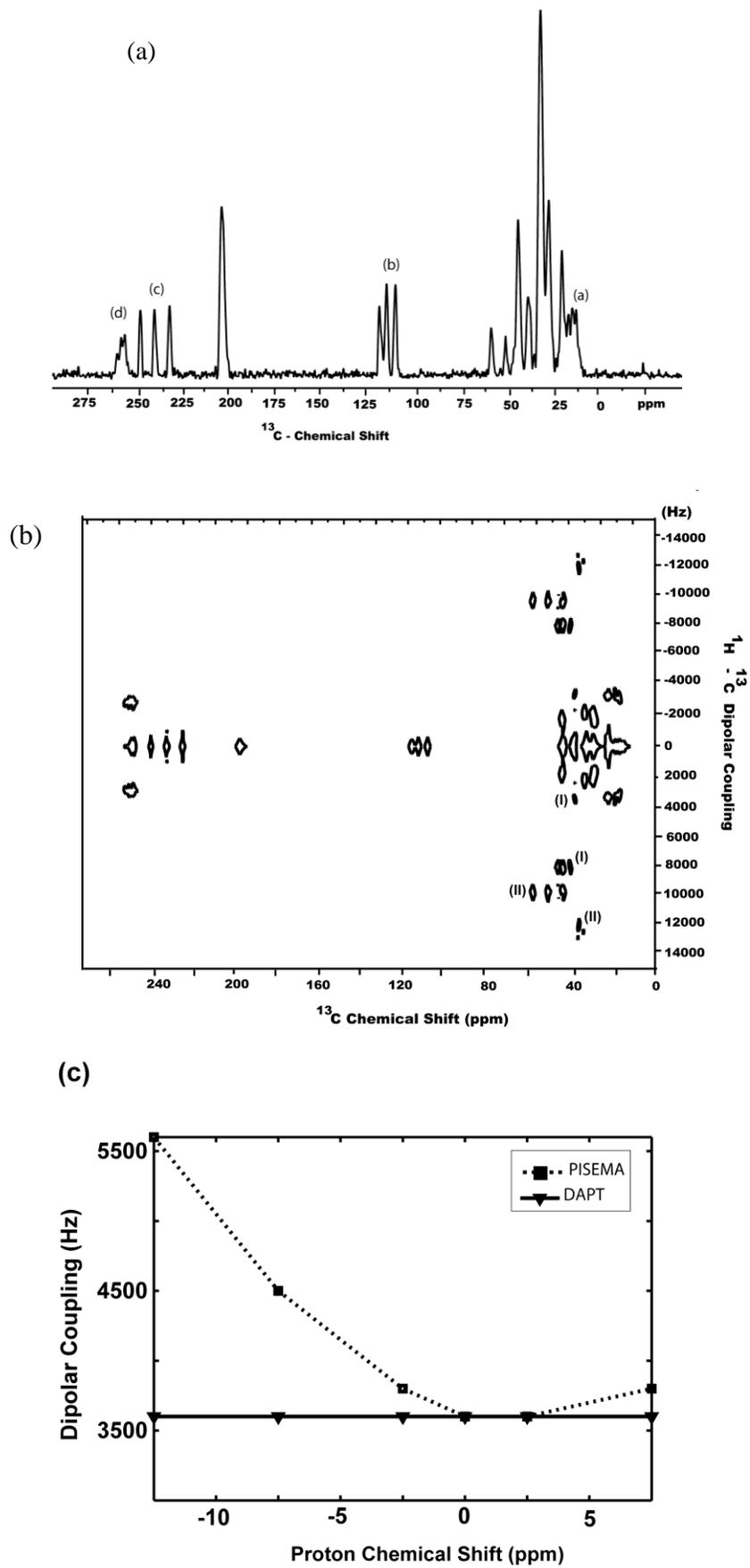

(a)

(b)

(c)

**Fig.4**

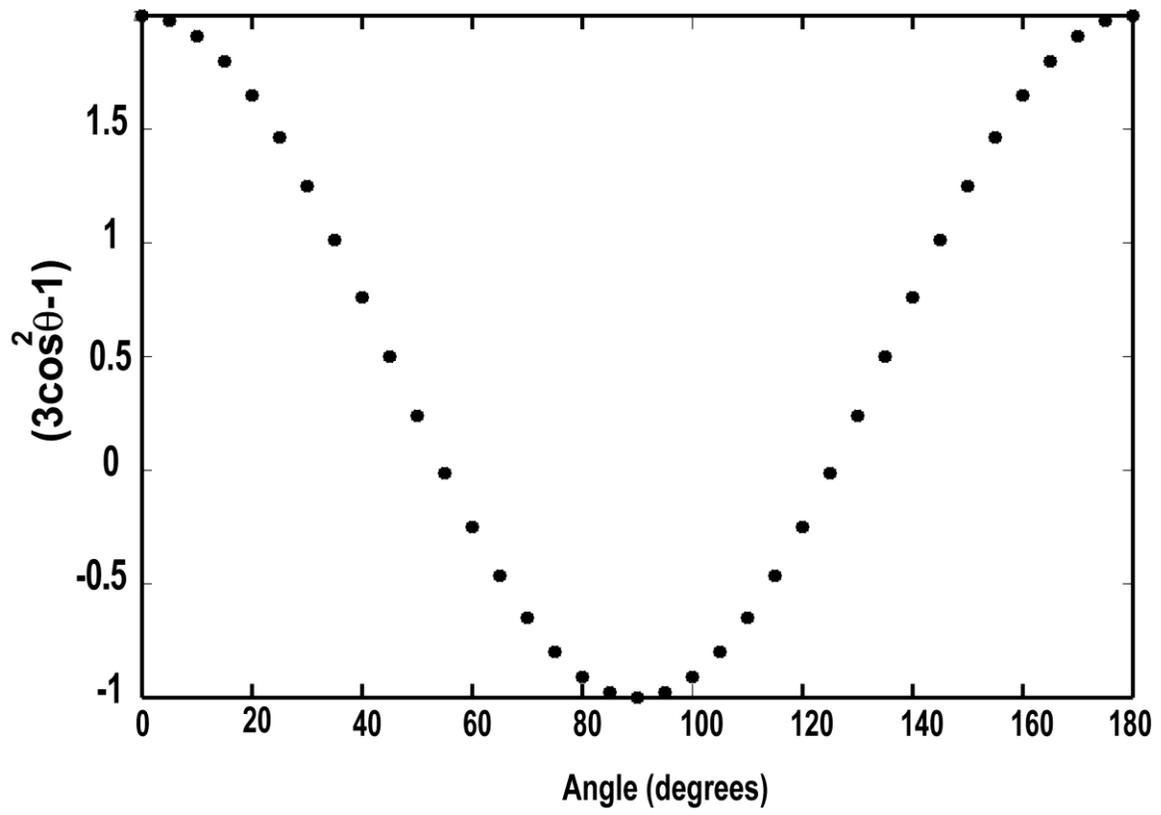

**Fig.5**

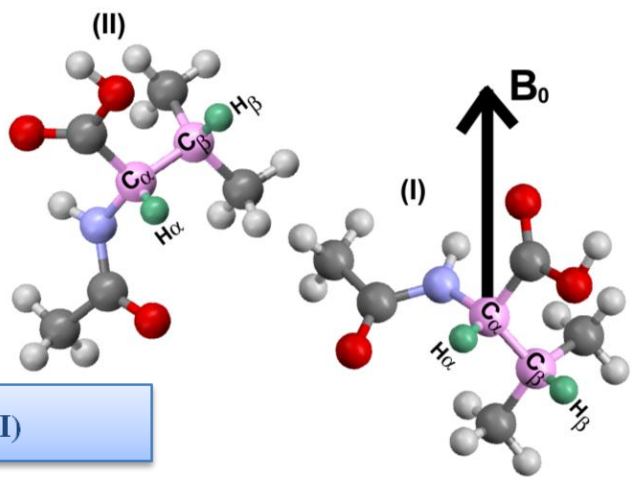

(II)

(I)

$B_0$

$H_\beta$

$C_\beta$

$C_\alpha$

$H_\alpha$

$C_\alpha$

$H_\alpha$

$C_\beta$

$H_\beta$

---

2D DAPT Spectrum : $+D_{IS}^\alpha(I)$ : Calculate $\theta^\alpha_+(I)$

Crystal coordinates : Change $C^\alpha(I)$ coordinate to ( 0 0 0) and fix $B_0$ at (0 0 1)

Crystal coordinates : Position $C^\alpha\text{-}^1H(I)$ by $\theta^\alpha_+(I)$ wrt to $B_0$ : Rodrigues' formula

Calculate the new coordinates of $C^\beta\text{-}^1H(I)$ : Calculate $\theta^\beta_+(I)$ and $D_{IS}^\beta(I)$

2D Spectrum: Is it matching with the Expt. values

**N**

2D DAPT : Consider $-D_{IS}^\alpha(I)$ : Calculate $\theta^\alpha_-(I)$

Crystal coordinates : Rotate $C^\alpha\text{-}^1H(I)$ by $\theta^\alpha_-(I)$ wrt to $B_0$ : Rodrigues' formula

Calculate the new coordinates of $C^\beta\text{-}^1H(I)$ : Calculate $\theta^\beta_-(I)$ and $D_{IS}^\beta(I)$

**Y**

Transform old coordinates of Molecule – II to new coordinate system.

Is it matching with Exp.?

**N**

Calculate $C^\alpha\text{-}^1H(II)$ and $C^\beta\text{-}^1H(II)$ And compare with 2D spectrum

Repeat the protocol from coordinate transformation onwards by swapping the molecule labels

**Y**

$C^\alpha\text{-}^1H$ and its corresponding $C^\beta\text{-}^1H$ identified

19